\documentstyle[amssymb,aps,12pt]{revtex}
\begin{document}
\draft
\author{Sergio De Filippo}
\author{Filippo Maimone}
\address{Dipartimento di Fisica ''E.R. Caianiello'', Universit\`{a} di Salerno\\
Via Allende I-84081 Baronissi (SA) ITALY\\
Tel: +39 089 965 229, Fax: +39 089 965 275, e-mail: defilippo@sa.infn.it\\
and \\
Unit\`{a} I.N.F.M., I.N.F.N. Salerno}
\date{\today}
\title{Entropic localization in non-unitary Newtonian gravity.}
\maketitle

\begin{abstract}
The localizing properties and the entropy production of the Newtonian limit
of a nonunitary version of fourth order gravity are analyzed. It is argued
that pure highly unlocalized states of the center of mass motion of
macroscopic bodies rapidly evolve into unlocalized ensembles of highly
localized states. The localization time and the final entropy are estimated.
\end{abstract}

Dealing with the quantum limits of the second law, a strongly related,
somehow even unavoidable, issue is that of its quantum foundations: ''...in
order to gain a better understanding of the degrees of freedom responsible
for black hole entropy, it will be necessary to achieve a deeper
understanding of the notion of entropy itself. Even in flat space-time,
there is far from universal agreement as to the meaning of entropy --
particularly in quantum theory -- and as to the nature of the second law of
thermodynamics''\cite{wald}. A way out of the subjective and vaguely defined
procedure of coarse graining is the assumption that even the evolution of a
closed system is affected by a fundamental nonunitarity, as suggested by
black hole formation and evaporation\cite{hawking}. On the other hand the
possibility that gravity may lead to nonunitary evolution was invoked also
with reference to the measurement problem and the transition to classicality 
\cite{karolyhazy}.

The rationale for such an assumption should finally be found in a future
full theory of quantum gravity. In the meantime in some recent papers\cite
{defilippo,defmaim} a specific non-Markovian nonunitary model for Newtonian
gravity, without any free parameter, was derived as the nonrelativistic
limit of a nonunitary version of fourth order gravity.

While for the non-Markovian nonunitary models considered by Unruh and Wald
the basic idea is to have the given system interacting with a ''hidden
system'' with ''no energy of its own and therefore... not... available as
either a net source or a sink of energy'' \cite{unruh}, in the present model
energy conservation is granted by the ''hidden system'' being a copy of the
physical system, coupled to it only by gravity, and constrained to be in its
same state and then to have its same energy. The unitary dynamics and the
states referred to the doubled operator algebra are what we call
respectively meta-dynamics and meta-states, while, by tracing out the hidden
degrees of freedom, we get the non-unitary dynamics of the physical states.
Pure physical states correspond then to meta-states without entanglement
between physical and hidden degrees of freedom.

In particular it has been shown that, while reproducing the classical
aspects of the Newtonian interaction, the model gives rise to a threshold,
which for ordinary condensed matter densities corresponds to $\sim 10^{11}$
proton masses, above which self-localized center of mass wave functions
exist. Moreover an initial localized pure state undergoes an entropic
spreading, namely it evolves (very slowly in time ) into an unlocalized
ensemble of localized states. That is consistent with the expectation that
(self-)gravity may produce a growing entropy in a genuinely isolated system,
as suggested by black hole physics.

We give here a concise definition of the model. Let $H[\psi ^{\dagger },\psi
]$ be the non-relativistic Hamiltonian of a finite number of particle
species, like electrons, nuclei, ions, atoms and/or molecules including also
the halved gravitational interaction, where $\psi ^{\dagger },\psi $ denote
the whole set $\psi _{j}^{\dagger }(x),\psi _{j}(x)$ of
creation-annihilation operators, i.e. one couple per particle species and
spin component. $H[\psi ^{\dagger },\psi ]$\ includes the usual
electromagnetic interactions accounted for in atomic, molecular and
condensed-matter physics. To incorporate that part of gravitational
interactions responsible for non-unitarity, one has to introduce
complementary creation-annihilation operators $\widetilde{\psi }%
_{j}^{\dagger }(x),\widetilde{\psi }_{j}(x)$ and the overall
(meta-)Hamiltonian 
\begin{equation}
H_{G}=H[\psi ^{\dagger },\psi ]+H[\widetilde{\psi }^{\dagger },\widetilde{%
\psi }]-\frac{G}{2}\sum_{j,k}m_{j}m_{k}\int dxdy\frac{\psi _{j}^{\dagger
}(x)\psi _{j}(x)\widetilde{\psi }_{k}^{\dagger }(y)\widetilde{\psi }_{k}(y)}{%
|x-y|}  \label{meta-hamiltonian}
\end{equation}
acting on the product $F_{\psi }\otimes F_{\widetilde{\psi }}$ of the Fock
spaces of the $\psi $ and $\widetilde{\psi }$ operators, where $m_{i}$ is
the mass of the $i$-th particle species and $G$ is the gravitational
constant. The $\widetilde{\psi }$ operators obey the same statistics as the
corresponding operators $\psi $, while $[\psi ,\widetilde{\psi }]_{-}=[\psi ,%
\widetilde{\psi }^{\dagger }]_{-}=0$.

The meta-particle state space $S$ is the subspace of $F_{\psi }\otimes F_{%
\widetilde{\psi }}$ including the meta-states obtained from the vacuum $%
\left| \left| 0\right\rangle \right\rangle =\left| 0\right\rangle _{\psi
}\otimes \left| 0\right\rangle _{\widetilde{\psi }}$ by applying operators
built in terms of the products $\psi _{j}^{\dagger }(x)\widetilde{\psi }%
_{j}^{\dagger }(y)$ and symmetrical with respect to the interchange $\psi
^{\dagger }\leftrightarrow \widetilde{\psi }^{\dagger }$, which, then, have
the same number of $\psi $ (physical) and $\widetilde{\psi }$ (hidden)
meta-particles of each species. As for the observable algebra, since
constrained meta-states cannot distinguish between physical and hidden
operators, it is identified with the physical operator algebra. In view of
this, expectation values can be evaluated by preliminarily tracing out the $%
\widetilde{\psi }$ operators. In particular, for instance, the most general
meta-state corresponding to one particle states is represented by 
\begin{equation}
\left| \left| f\right\rangle \right\rangle =\int dx\int dyf(x,y)\psi
_{j}^{\dagger }(x)\widetilde{\psi }_{j}^{\dagger }(y)\left| 0\right\rangle
,\;\;f(x,y)=f(y,x).  \label{f}
\end{equation}
This is a consistent definition since $H_{G}$\ generates a group of
(unitary) endomorphisms of $S$.

Consider now a uniform matter ball of mass $M$ and radius $R$. Within the
model the Schroedinger equation for the meta-state wave function $\Xi
(X,Y,t) $ is given by 
\begin{equation}
i\hslash \frac{\partial \Xi }{\partial t}=\left[ -\frac{\hslash ^{2}}{2M}%
(\nabla _{X}^{2}+\nabla _{Y}^{2})+V(\left| X-Y\right| )\right] \Xi \equiv 
{\it H}_{G}\Xi  \label{schroedinger}
\end{equation}
where $X$ and $Y$ respectively denote the position of the center of mass of
the physical body and of its hidden partner, while $V$ is the (halved)
gravitational mutual potential energy of the two interpenetrating
meta-bodies, whose explicit form can be found in Ref. \cite{defmaimrob}.

In particular in Ref. \cite{defmaimrob} it was checked numerically that a
slightly unlocalized pure state for such a matter ball just above the mass
threshold evolves slowly into a mixed state with a small nonvanishing von
Neumann entropy. Since such a state has approximately vanishing coherences
for space points farther than the width of a highly localized state, it was
argued that it can be obtained by tracing out the hidden degrees of freedom
from a linear combination of highly localized bound metastates. In the
present paper we want to show that a highly unlocalized pure state may
evolve into an entropic unlocalized ensemble of highly localized states in a
short time, even though meta-energy conservation prevents the formation of
highly localized bound metastates. In order to do that, consider an initial
Gaussian wave function $\Psi (X)\propto \exp [-X^{2}/\Lambda _{0}^{2}]$,
corresponding to an unentangled meta-wavefunction 
\begin{eqnarray}
\Xi \left( X,Y;t_{0}\right) &=&\Psi (X)\Psi (Y)\propto \exp
[-(X-Y)^{2}/(2\Lambda _{0}^{2})]\exp [-(X+Y)^{2}/(2\Lambda _{0}^{2})]\equiv 
\nonumber \\
\Phi (X-Y,t &=&0)\Theta (X+Y,t=0),
\end{eqnarray}
and assume that $\Lambda _{0}$\ is large enough to make (the expectation of)
kinetic meta-energy irrelevant with respect to potential meta-energy, which,
assuming $\Lambda _{0}$ $\gg R$, can be identified with the Newton energy
for point particles. On the other hand, by the quantum virial theorem,
(assuming, as it is natural, that the metastate can be well approximated by
a linear combination including bound metastates only) we have for the time
average $\overline{\left\langle K\right\rangle }$ of the expectation $%
\left\langle K\right\rangle $ of the kinetic energy $K$ of the relative
motion: 
\begin{equation}
\overline{\left\langle K\right\rangle }\simeq -\left\langle {\it H}%
_{G}\right\rangle \simeq \frac{GM^{2}}{\Lambda _{0}}.  \label{approximation}
\end{equation}
The corresponding wave function $\Phi (X-Y,t)$ at a generic instant of time $%
t$ is expected to give an expectation for $K$ approximately coinciding with
its time average, the initial value being exceptional. This corresponds to a
typical length for phase variations of $\Phi (X-Y,t)$ given by 
\begin{equation}
\Lambda \sim \frac{\hslash }{\sqrt{2M\left\langle K\right\rangle }}\sim
\hslash \sqrt{\frac{\Lambda _{0}}{GM^{3}}},
\end{equation}
by which, if we consider the physical state 
\begin{equation}
\rho (X,X^{\prime },t)=\int dY\Phi (X-Y,t)\Theta (X+Y,t)\Phi ^{\ast
}(X^{\prime }-Y,t)\Theta ^{\ast }(X^{\prime }+Y,t),
\end{equation}
we have that (while the factors $\Theta (X+Y,t)$ and $\Theta ^{\ast
}(X^{\prime }+Y,t)$ are slowly varying in space, since, for $M$ and $\Lambda
_{0}$ large enough they essentially coincide with their value at $t=0$,
apart from an exceedingly slow spreading) the factors describing the
relative motion have very rapid phase variations on the typical space scale $%
\Lambda $, this giving rise to cancellations except for $\left| X-X^{\prime
}\right| \lesssim \Lambda $, when the product $\Phi (X-Y,t)\Phi ^{\ast
}(X^{\prime }-Y,t)$ is approximately real and positive.

We can conclude that, after a typical localization time $\tau _{l}\sim
\hslash \Lambda _{0}/(GM^{2})$, necessary for the Newton interaction to
affect the state, the physical state $\rho (X,X^{\prime },t)$ has vanishing
coherences for $\left| X-X^{\prime }\right| \gg \Lambda $, by which it is
natural to assume that it can be represented as an ensemble of localized
states with a localization length $\sim $ $\Lambda $. The number $N$ of
these states can be approximately evaluated as 
\begin{equation}
N\sim \frac{\Lambda _{0}^{3}}{\Lambda ^{3}}=\frac{\Lambda
_{0}^{3/2}G^{3/2}M^{9/2}}{\hslash ^{3}},
\end{equation}
which, in the assumption that $N$ is very large, allows to estimate the
entropy as that of an ensemble of equiprobable states, namely as $S\simeq
K_{B}\ln N$.

To give a numerical example, consider the case $M=10^{-9}g$, $R=10^{-3}cm$, $%
\Lambda _{0}=10^{-1}cm$, which gives $\Lambda \sim 10^{-10.5}cm$, $N\sim
10^{28.5}$, $S\simeq K_{B}(\ln 10)28.5$, $\tau _{l}\sim 10^{-3}\sec $. It is
immediate to check that the approximate equalities in Eq. (\ref
{approximation}), depending on $\Lambda _{0}$ being large enough, are in
such a case practically exact. Furthermore an explicit evaluation gives $%
\left\langle {\it H}_{G}\right\rangle =-3.31\cdot 10^{-24}erg$, $\sqrt{%
\left\langle {\it H}_{G}^{2}\right\rangle -\left\langle {\it H}%
_{G}\right\rangle ^{2}}=3.09\cdot 10^{-24}erg$, which is consistent with our
assumption that the meta-state has a very small projection on the subspace
of scattering metastates. Finally the spreading time of the center of mass
motion is $\sim 10^{16}$ sec.

Some comments are in order to the present result. First, with reference to
the quantum foundations of the second law, it shows only that in principle
the present model can give rise to a growth of the von Neumann entropy even
for a closed system. Of course this is possible because we start from a pure
physical state, which within the model is a highly unlikely state for the
center of mass motion of a macroscopic body. It is the analogue, for the
entropy of the inner degrees of freedom of an ordinary matter system, of a
state very far from equilibrium, while the localization process plays the
role of the evolution towards equilibrium. To be specific, in the numerical
example the principal quantum numbers of the hydrogen-like bound metastates
corresponding to the average energy are $\sim 10^{10}$. This means that so
many stationary metastates enter in the expression of the time dependent
metastate that we have a substantial ergodicity with extremely long
recurrence times. However possible in principle, the explicit evaluation,
within the model, of the entropy growth for a realistic matter system is a
rather hard task, which led us to consider a highly idealized initial state
just of the center of mass motion of a matter ball.

As to the measurement problem, the initial pure state can be seen as a
delocalized pointer state, while the final mixed one is the result of a
dynamical wave function reduction. (The inclusion of a measured microscopic
system would not change the picture \cite{defmaim}.)

The most attractive feature of the model is precisely this possibility to
address simultaneously, by a nonunitary version of Newtonian gravity without
any free parameter, classically equivalent to the standard one, two
unsettled questions: the quantum foundations of the second law and the
transition to classicality. Another feature of the nonrelativistic model is
the absence of any obstruction to its special-relativistic extension, at
variance with other localization models \cite{defmaim2}.

\end{document}